\documentclass[
]{ceurart}

\usepackage[frozencache,cachedir=.]{minted}
\setminted{breaklines=true}

\newcommand{\note}[1]{}
\begin{document}

\copyrightyear{2025}
\copyrightclause{Use permitted under Creative Commons License Attribution 4.0 International (CC BY 4.0).}

\conference{Joint Proceedings of the ACM IUI Workshops 2025, March 24-27, 2025, Cagliari, Italy.}

\title{Talking Back -- human input and explanations to interactive AI systems}


\author[1,2]{Alan Dix}[%
orcid=0000-0002-5242-7693,
email=alan@hcibook.com,
url=https://alandix.com/,
]

\author[3]{Tommaso Turchi}[orcid=0000-0001-6826-9688,email=tommaso.turchi@unipi.it]

\author[2]{Ben Wilson}[orcid=0009-0004-5663-5854,email=b.j.m.wilson@swansea.ac.uk]

\author[3]{Anna Monreale}[orcid=0000-0001-8541-0284,email=anna.monreale@unipi.it]

\author[2]{Matt Roach}[orcid=000-0002-1486-5537,email=m.j.roach@swansea.ac.uk]


\address[1]{Cardiff Metropolitan University, Wales, UK}
\address[2]{Computational Foundry, Swansea University, Wales, UK}
\address[3]{Department of Computer Science, University of Pisa, Pisa, Italy}


\begin{abstract}
  While XAI focuses on providing AI explanations to humans, can the reverse -- humans explaining their judgments to AI -- foster richer, synergistic human-AI systems? This paper explores various forms of human inputs to AI and examines how human explanations can guide machine learning models toward automated judgments and explanations that align more closely with human concepts.
\end{abstract}


\maketitle
























   

\section{Introduction}

It is now well accepted that it is important for AI systems to offer explanations to users \cite{10.1145/3313831.3376590}; indeed this is embedded in the EU GDPR legislation \cite{ceu2016gdpr}.  This paper poses the converse question, ``what if users offered explanations of their decisions to an AI system''.

One of the reasons for transparency and XAI (explainable AI) is to surface potential bias as first noted by one of the authors in the early 1990s \cite{dix1992human}.  This is equally important for humans, especially when their input is being used to train an AI system, which might otherwise embody the trainer's prejudices.

However, human explanations offer other advantages, allowing machine learning (ML) that is more aligned with human reasoning, and can itself be explained in more human-like ways.

In the rest of this paper, we will first look more broadly at the ways humans currently engage with AI systems in terms of the kinds of inputs they provide (Section~\ref{sc:human-input}).  We then look at ways in which human explanations might be provided at the user interface level (Section~\ref{sc:explain}).  This will create a series of examples of different kinds of user explanations, which raises algorithmic (Section~\ref{sc:algorithm}), architectural (Section~\ref{sc:arch}), and even social challenges (Section~\ref{sc:social}).

Once one thinks in terms of human explanations, it is possible to see elements of this in previous work, not least classic knowledge elicitation techniques \cite{shadbolt2015knowledge}.  However, it appears to be a largely unexplored area, and we hope to encourage others to explore it with us.

\section{Human inputs to AI}
\label{sc:human-input}

\begin{figure}
  \centering
  \includegraphics[width=0.5\linewidth]{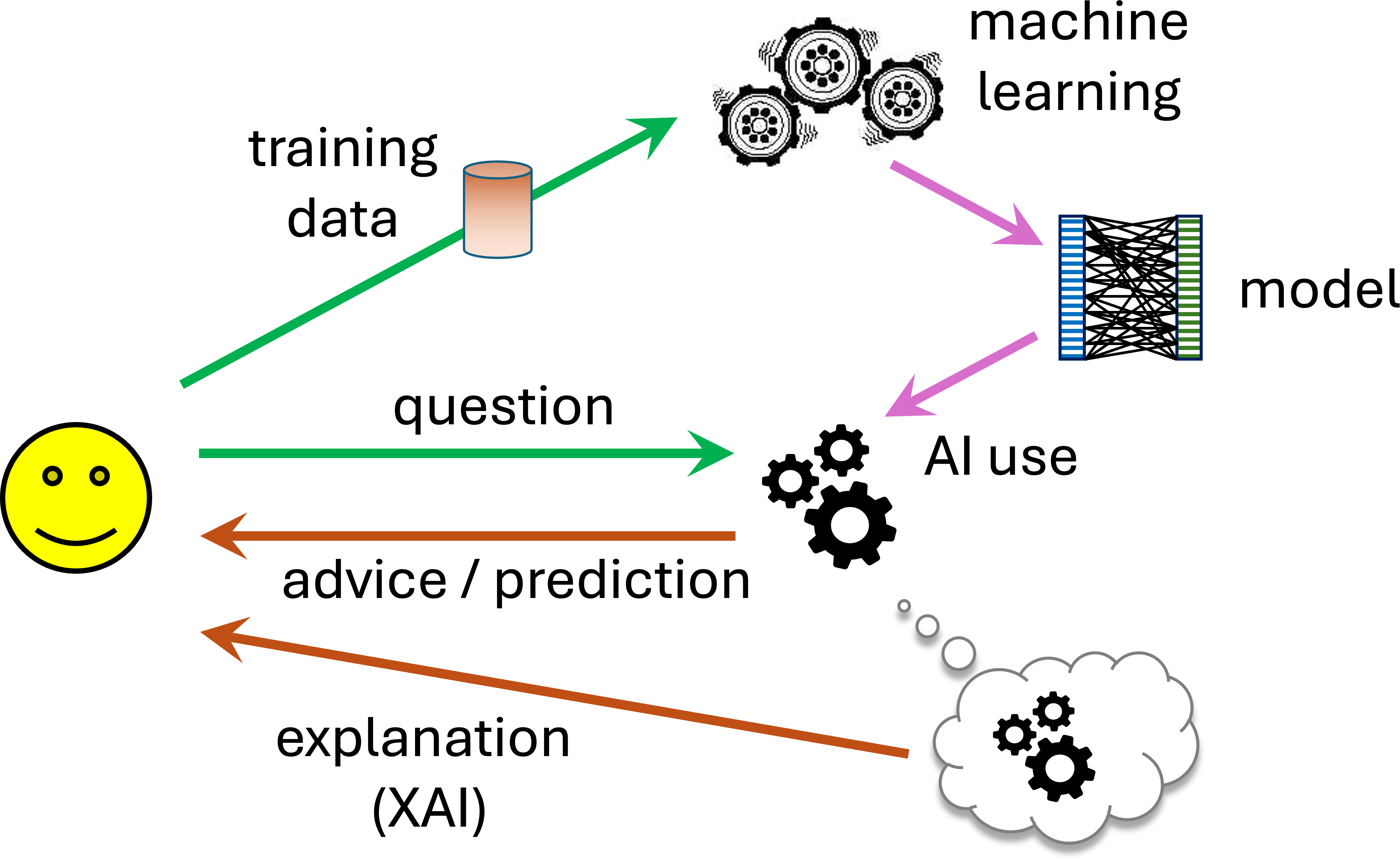}
  \caption{Common communications between human user and AI.}
  \label{fg:common-hai}
\end{figure}

With the exception of chatbots, humans and AI typically take very different kinds of roles during an interaction.  Figure~\ref{fg:common-hai} show some common ways in which they communicate.  During learning a human expert may provide labeled training examples, which are used to create a model such as a neural network, decision tree or rule set.  This model is then used in practice when the user makes some sort of query, perhaps presenting an unseen data item to classify, whereupon the AI system provides a predicted classification or other form of advice or answer.  Often also an XAI system will create some form of explanation, that to some extent `explains' the AI system's response.  Most commonly this is a \emph{local explanation} that explains the classification or decision about a particular data item and those close to it, rather than a \emph{global explanation} of the model as a whole. 

\begin{figure}
  \centering
  \includegraphics[width=0.7\linewidth]{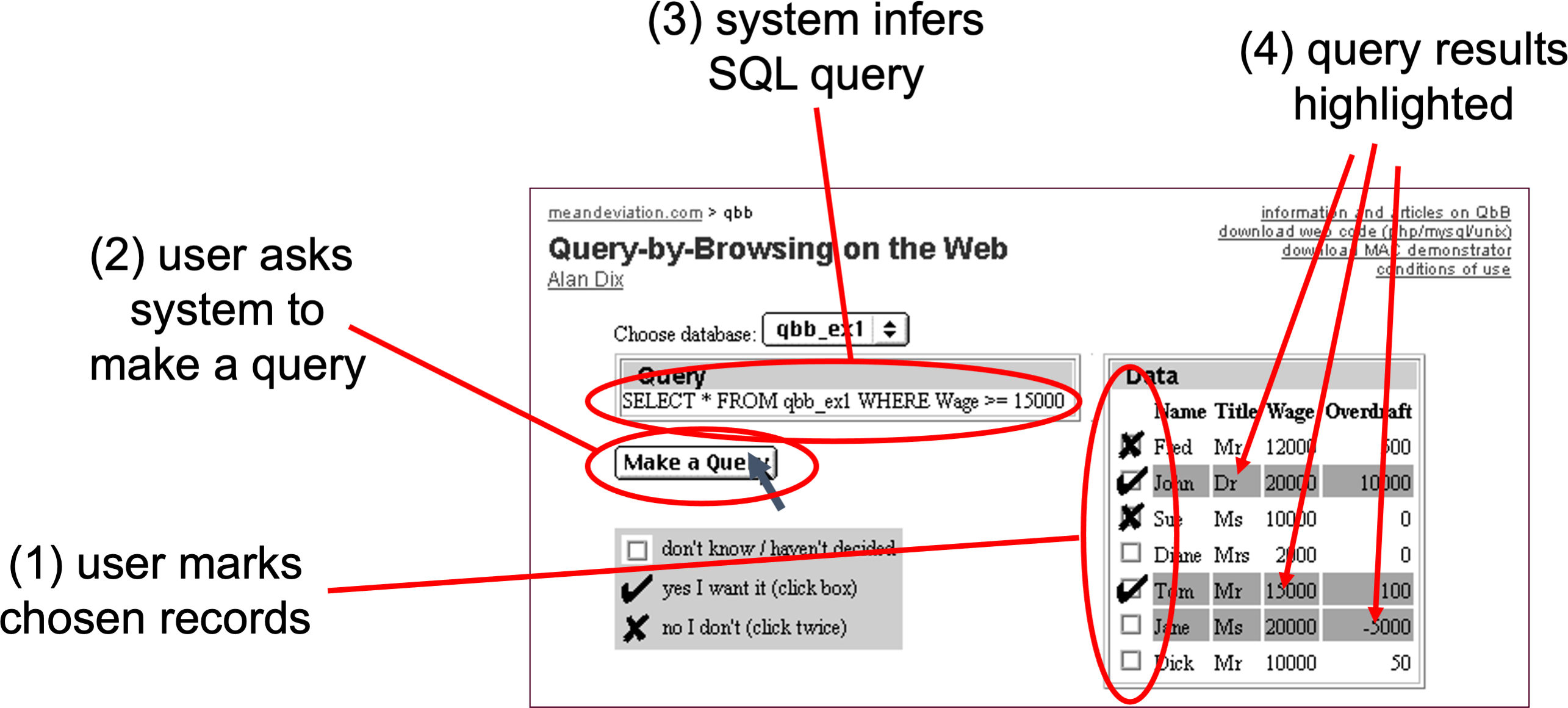}
  \caption{Query-by-Browsing -- how it works \protect\cite{dix1994query}.}
  \label{fg:qbb-hiw}
\end{figure}

If we focus on the human inputs, we can identify a number of common ways that knowledge is passed to the AI:

\begin{description}

\item[example data] -- This may be part of a batch exercise during training, or may be more interactive. 
 For example, in Query-by-Browsing (QbB; see Fig.~\ref{fg:qbb-hiw}), the user looks through a database listing selecting some records of interest and rejecting others; after a while the system infers an SQL query \cite{dix1994query}.  However, if this query is not correct, the user can simply provide more positive or negative examples of data records, leading to a revised system SQL query.

\item[labeling] -- Training data is often labeled, but users may also be explicitly asked to label data items for which the algorithm has low confidence, or to help verify learnt rules. In particular during interactive learning \cite{ware2001interactive,fails2003interactive,dudley2018review}, the expert presents training data, but the system may challenge these with counterfactuals, leading to a form of `conversation' where each utterance is a labeled data item.

\item[relevance/quality feedback]  \note{Tom's notes say `RHLF'} -- Often users are asked to give some form of explicit star rating or thumbs up, which is used to incrementally improve relevance models.  This can be thought of as a form of post-hoc labeling.

\item[implicit sensing] -- The training data and/or labeling may be captured implicitly from one or more humans, for example using physical sensors for activity recognition, click-through data for search optimisation, or purchase/viewing history for recommender systems.  This form of data gathering can be made more effective through careful user interface design, such as epistemic interaction \cite{dix2024epistemic}.

\end{description}

As well as these common forms of human information provision, there are some less common ones:

\begin{description}

\item[confirming/rejecting system rules] -- In some variants of Query-by-Browsing when the system generates a rule, the user can reject it and then the system would search for an alternative rule.

\item[editing/amending system rules] -- When the model used by the AI is a transparent set of rules, it may be possible for the user to tinker with them. One variant of QbB used a genetic algorithm to generate conjunctive rules displayed in a query-by-example-style tableau \cite{dix1998interactive}.  In this case the user could accept a rule, but also edit some of the parts of it, for example, changing the threshold in a `greater than' clause.  Similarly, the paper that first introduced the term `interactive machine learning', described a system where the user and machine learning system worked collaboratively to create a decision tree \cite{ware2001interactive}.

\item[confirmation of explanations] -- When the model is more obscure, local or global explanations may be given.  In these cases the user may accept, reject or rate an explanation.  This does not mean not so much that the explanation is not correct in the sense of matching the model's behaviour (one presumes that the XAI engine ensures this), but more whether it is comprehensible or useful.  In the case of an expert this might be saying ``I agree with the model's output, but this is not the right reason for it'', for the novice end user more ``yes that was helpful''. This can be used as a form of relevance/quality feedback for the XAI engine.

\item[editing/amending explanations] -- FRANK \cite{mazzoni2024frank} uses a `skeptical learning' approach \cite{monreale2024cognition}, which allows the user to reject or modify an explanation.  Similarly \cite{teso2019explanatory} ask a human annotator to critique a LIME-style explanation and identify ``the components that have been wrongly identified by the explanation as relevant''.

``\textit{Explanations can be extracted using feature-based methods. At this point, the annotator supplies a label for the instance and, optionally, corrective feedback on the explanation. This means that the user
can, for instance, indicate what input variables the machine is wrongly relying on for making
its prediction.}'' \cite[p.17]{monreale2024cognition}

\end{description}

Note how the ability to edit rules and explanations can be seen as a form of \emph{equal opportunity}  \cite{runciman86equal}, the design principle that suggests that when possible rather than having system output and user input separately, each partner in the dialogue should have access.  For example, the system might be able to add defaults/suggestions in user inputs fields.  Similarly, where possible, the user can be allowed to edit output values, with this updating appropriate internal data, or at very least be able to reuse the output values.

\begin{figure}
  \centering
  \includegraphics[width=0.5\linewidth]{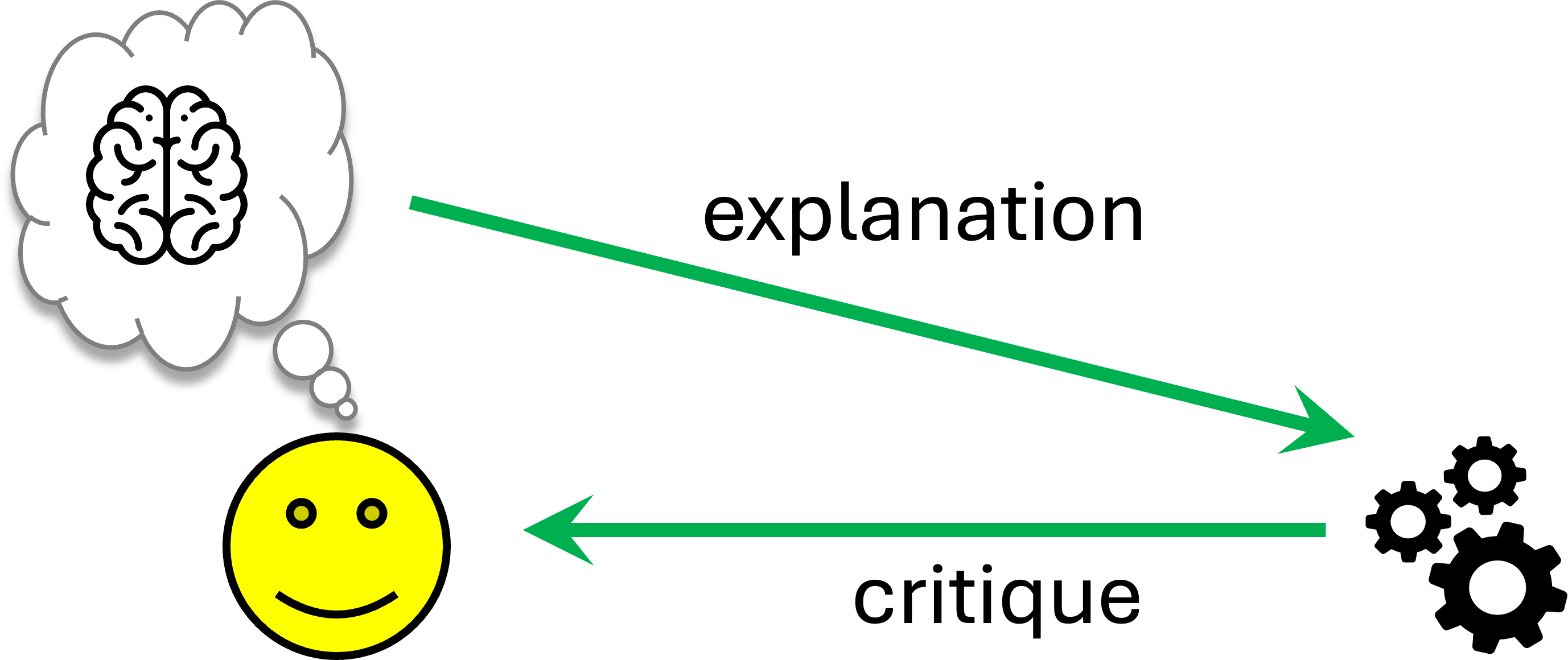}
  \caption{Some less common communication types between human user and AI.}
  \label{fg:uncommon-hai}
\end{figure}

However, one can apply this principle a little more deeply, looking at the kinds of ways the user and AI system present information and take input from each other and ask if there are parallels the other way round.  Figure~\ref{fg:uncommon-hai} shows two ways in which this might extend the idea of who does what:

\begin{description}

\item[AI as critic] -- If the user can approve or edit system results and explanations, could the AI system also be able to take on the role of critic?  In fact, existing systems can be seen in this light.  Skeptical learning does ask the user to think again if the user's classifications appear to contradict previous training examples \cite{mazzoni2024frank}, this can be seen as an AI parallel of the relevance feedback or confirmation of system outputs.  At a  richer level, DebunkBot, a chatbot that provides evidence to challenge misinformation, has been found to reduce belief in conspiracy theories \cite{costello2024durably}. \note{plus Peter Daish's system}

\item[user explanations] -- If systems are asked to produce explanations of their reasoning, what if users could do the same?  For example, when an expert labels a candidate's CV as being a good fit for an academic job, they could say it is because, ``\texttt{has\_PhD} is \texttt{Yes} and \texttt{number\_of\_publications} is large''.  Such information could be used both to guide ML systems to create models that correspond more closely to the human's decision making and also to provide a vocabulary whereby the system can create better explanations of its own operation.

\end{description}

Both of these open up interesting research questions, but in this paper we will focus on the latter.

\note{Something about vocabulary: data items, input features, classifications, yes/no, rating, human-new-feature, free text, XAI-new-features }

\section{Human explanations}
\label{sc:explain}

We commonly offer explanations of our own views or decisions when interacting with other humans.  Just as with XAI, this may include global explanations, offering some sort of overarching rule, or more often a local explanation, justifying a specific decision.  Explanations of this sort will, by their nature, be given by someone with a level of domain expertise; however the degree of expertise may vary.  In some cases the explanation may therefore be treated as ground truth, whereas in others more advisory.

In many cases these explanations may be incomplete, and typically only operate for `examples like this one'.  For example, during a job selection process, one might be asked, ``why this candidate'';  a possible answer could be ``because they have a PhD and high publication rate''.  However, on a different occasion the answer might be ``because of extensive industrial experience and large grant income''.  We understand that these are local criteria, although one could also be asked to justify the apparent contradiction, or think about difficult cases where the two would lead to different outcomes.

In this section we will look at how such human explanations might be included in different kinds of user interactions with AI.

\subsection{Feature constraints in QbB}
\label{sc:ui-qbb}

In Query-by-Browsing \cite{dix1994query} the user's interaction is to simply add or remove examples, as shown in Figure~\ref{fg:qbb-hiw}.  The user is saying ``this is a wanted/unwanted row'', but not why.

\begin{figure}
  \centering
  \includegraphics[width=0.4\linewidth]{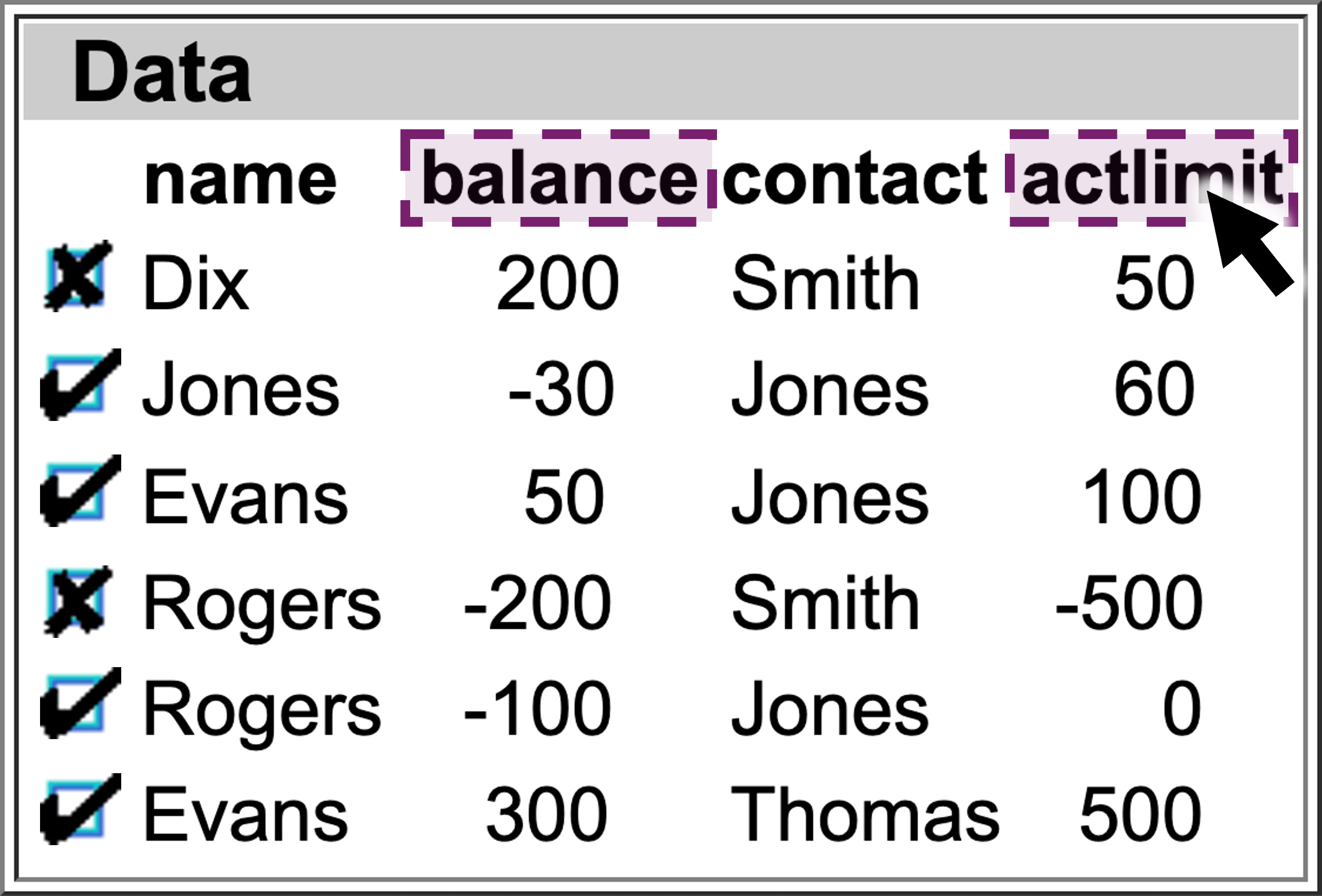}
  \caption{QbB -- global feature selection.}
  \label{fg:qbb-global}
\end{figure}

One could easily add the ability to have a degree of global `explanation'.  Figure~\ref{fg:qbb-global} shows how this might be done; as well as selecting the wanted/unwanted rows, the user also selects one or more columns.  This is saying, ``these are the rows I want and these are the columns I'm using to make my judgment''.  Arguably this is not so much an explanation as simply specifying to the algorithm which input features to use.

\begin{figure}
  \centering
  \includegraphics[width=0.5\linewidth]{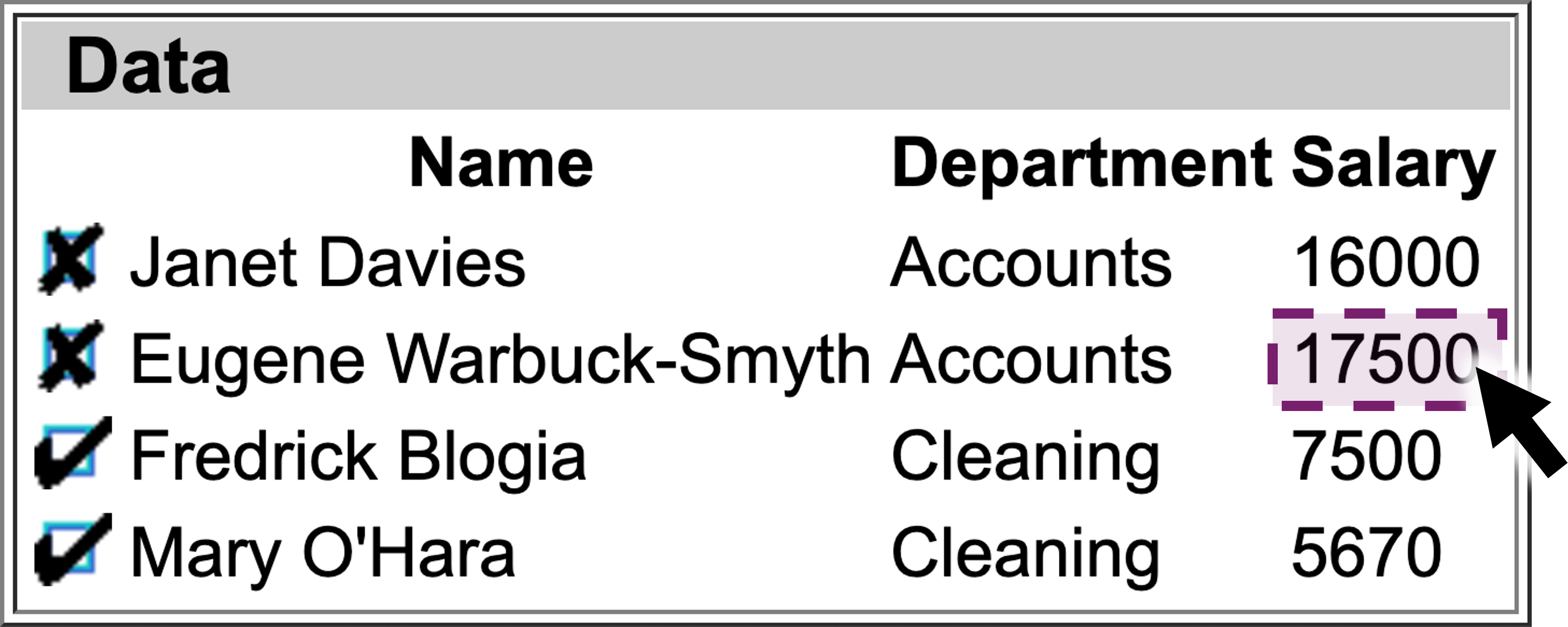} \\
  \vspace{1em}
  \includegraphics[width=0.5\linewidth]{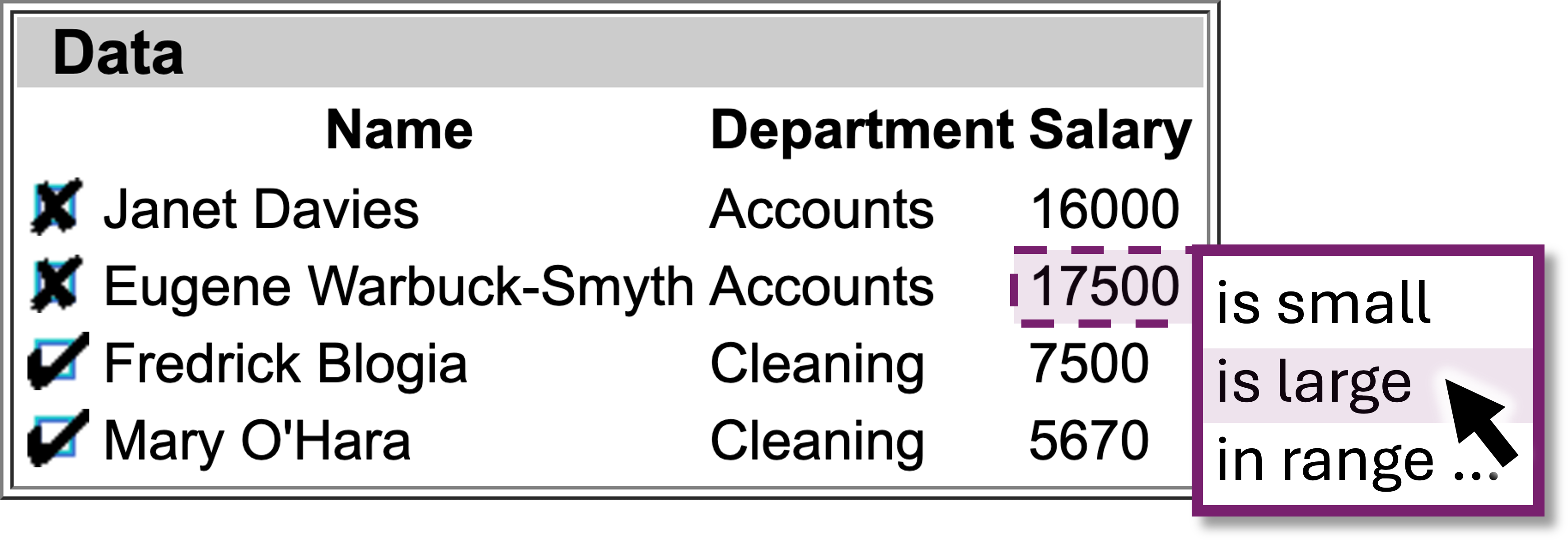}
  \caption{QbB -- local feature explanations: (above) select feature(s); (below) specify local rule.}
  \label{fg:qbb-local}
\end{figure}

It would also be possible for the user to specify columns for a specific row (see Fig.~\ref{fg:qbb-local}, above), which would be mean, ``I chose this row because of this column value''.  This feels more like a true explanation.

The user could be more precise specifying what it is about the feature that made it useful to make the discrimination (see Fig.~\ref{fg:qbb-local}, below).  This could be simply giving the general direction, `salary is large', or an exact rule such as, `salary greater than 10,000'. 

These user explanations can be regarded as constraints on the algorithm, saying that when the system makes a decision about this case, the specified features or rules should be part of the process that gave rise to it.  This should mean that the system's decision model is more closely aligned with that of the user.

It is easy to only provide specified features in the training data for a machine learning algorithm -- that is, making use of the global `explanation'.  However, these local explanations may require more radical modifications to ML algorithms.

An ML research challenge is to tune the ML so that the user-provided criteria are guaranteed (or encouraged to be) part of the ML model.  For decision trees or rules determined by a genetic algorithm, this may be a fairly straightforward change to the fitness function  For DNNs this might be encouraged by adding the criteria explicitly into pinch-point layers \cite{wust2024Pix2CodeLearningCompose}, or creating clouds of generated data around the key examples classified by the criteria.

\subsection{Counterfactuals in skeptical learning}
\label{sc:ui-il}

The Frank system employs techniques described as `skeptical learning' \cite{mazzoni2024frank}.  The user provides training data, but the system challenges the user when new examples appear to contradict previous ones, effectively saying ``you've just said X is class C, but you said Y, which is very similar, is is class D''.  This forces the user to think again about the example.  As a result of this the user might retract, relabel or revise the new example or a previous one.

Could the user offer explanations, potentially local explanations (like local XAI) as part of this interaction?

\begin{enumerate}
\item User submits new example X – which the system determines potentially contradictory
\item System says “are you sure?”
\item User says “yes”
\item System says “but what about Y” and presents counterfactual that is very close to X, but with different class/diagnosis
\item User says “but in X property A  is much larger than in Y”
\item Maybe further rounds of interaction where the system finds another example Z where A is larger, eliciting further local explanations about a property B, etc ...
\end{enumerate}

Assuming the user does not retract X as part of this process, the system now needs to relearn including X, but also has properties A, B, ... that can and should be part of the local decision around X.

Note that, while the underlying ML algorithms are different, these explanations are like those suggested for Query-by-Browsing.  However, there the explanations were about a single item, whereas here the user explanation was in response to a counterfactual and is effectively making a distinction X is different from Y because of A.  

This raises further algorithmic challenges. It would be possible to treat the user explanation as two single item explanations: ``X because of A'' and ``Y because of not A'', but this overlooks key aspects of a differential explanation, not least that shared features between X and Y would not be mentioned.

\subsection{Feeding-back importance measures}
\label{sc:ui-importance}

\note{Shap-style levels of importance}

\note{Could note that while QbB demonstrates simple used/not used, this can be more continuous}

\note{Tom's notes say: at a higher level how would this work?}

\note{Some forms of explanations (e.g. Shap plots) include some form of importance measure or (image algorithms) hotspot visualisation.}

\note{Can similar techniques be used by the human, identifying the relative importance of different features in a training example for their decision about it?}

While Query-by-Browsing demonstrates the effectiveness of binary feature selection in database query contexts, other domains and situations may benefit from a more fine-grained approach to feature importance. Modern explainable AI techniques, particularly SHAP (SHapley Additive exPlanations) \cite{lundberg2017unified}, provide visualizations that show the relative contribution of different features to a model's prediction. This same paradigm could be reversed, allowing humans to indicate the relative importance of features in their own decision-making process.

For example, when classifying a medical diagnosis, rather than indicating a binary set of relevant symptoms, an expert could specify that elevated white blood cell count had a strong positive influence (0.8 on a normalized scale), while fever had a moderate positive influence (0.4), and patient history had a weak positive influence (0.2). This richer feedback more accurately captures the nuanced way humans weigh evidence in their decision-making process.

Such continuous importance measures could be captured through various interface mechanisms:
\begin{itemize}
    \item Direct manipulation of SHAP-style bar charts, where users can adjust the length and direction of bars representing feature importance;
    \item Interactive slider controls for each relevant feature, allowing precise specification of importance values;
    \item Visual annotation tools for marking regions of importance in image data, with intensity of markup indicating degree of importance;
    \item Relative ranking interfaces with automatic normalization to maintain consistent scales.
\end{itemize}

This approach provides a natural vocabulary for dialogue between human and AI systems, as both can express their reasoning in terms of feature importance distributions that can be directly compared and analyzed. It is particularly valuable in domains where feature interactions are complex and the relative importance of different factors varies significantly between cases.

The challenge then becomes how to incorporate these weighted importance measures into the learning process --- both for improving the model's predictions and for generating explanations that better align with human reasoning patterns. This alignment is particularly crucial in cases where the model and human may arrive at the same conclusion but through different reasoning paths, as indicated by divergent feature importance distributions. By enabling this richer form of feedback, we can work towards AI systems that not only learn from human decisions, but also from the reasoning process behind those decisions.

\subsection{User generated intermediate features}
 
\note{May be in bullet list, but not have full section}

\note{From Tom's notes: Example of model-based explanation feedback: higher level of causality wrt. SHAP or QbB, it not an explanation coming directly from one/more columns ...}

A user's explanation may include high-level terms or features that are not part of the input feature set, but derived from it.  For example, a doctor might say that an important aspect of their diagnosis is an ``irregular heart rhythm'', and label some data with this additional high-level feature, which is mid-way between the input data and final diagnostic classification. In some cases, the user may be able to identify which parts of the raw data contributed to this (e.g. an ECG trace).

Note we are referring here to derived features that the expert uses when seeing the same data as the AI.  This is different from times when the training expert introduces new vocabulary at the data capture stage, for example as part of a physical examination of a patient a doctor might say, ``the patient has a gray pallor''.  In the latter case the new vocabulary is an additional input space feature.

\subsection{Free text -- Chatbots and LLMs}
\note{Chatbot example: one of the responses
       => use this more \emph{because} ...
               => feed into memory \emph{Bondielli}}

While structured feedback mechanisms such as feature importance measures provide valuable input to AI systems, the emergence of Large Language Models (LLMs) opens up possibilities for more natural, free-text explanations from users. Modern LLMs often employ Reinforcement Learning from Human Feedback (RLHF) techniques \cite{ouyang2022traininglanguagemodelsfollow}, where they present multiple alternative responses and ask users to select the most appropriate one. While traditional RLHF typically only captures the binary or scalar preference between responses, this interaction pattern can be extended to capture not just which response users prefer, but also \emph{why} they prefer it.

For example, when an LLM presents two alternative explanations for a scientific concept, a teacher might select one and explain: ``This explanation uses more appropriate vocabulary for my middle school students'' or ``This version better connects to previous concepts we've covered in class.'' Such free-text rationales provide rich contextual information that goes beyond simple preference selection.  In particular, they may include the user explaining their critique of the system's explanation as well as the system decision.

This approach represents a significant departure from the methods discussed previously. Rather than working through structured interfaces, users can provide feedback in natural language, introducing new concepts and criteria not previously encoded in the system. The explanations can capture complex, context-dependent reasoning that might be difficult to express through feature importance or other structured mechanisms. Consider a medical professional explaining their preference for one diagnostic explanation over another: their rationale might encompass subtle interactions between patient history, current symptoms, and treatment context that would be challenging to capture through more structured feedback methods.

The primary challenge lies in effectively incorporating these free-text explanations into the system's memory and learning process. Future research could explore various approaches to extract structured knowledge from these explanations, identifying recurring themes, preferences, and reasoning patterns. This structured knowledge could then inform the system's response generation more systematically, potentially creating a feedback loop where human explanations continuously refine the model's understanding of different contexts and user needs.

This capability is particularly valuable in educational and advisory contexts, where the same information might need to be presented differently depending on the audience or situation. By accumulating user explanations about why certain responses work better in specific contexts, the system can build a more nuanced understanding of how to adapt its communication style and content appropriately.

\section{Algorithmic implications}
\label{sc:algorithm}

In the previous section we saw types of explanations which the user might present to the system:

\begin{description}

\item[hard feature constraints] -- that say that a feature must be included in the way that a data item is classified.  This may be precise $age>=18$, imprecise ``$age$ is large'' or unspecified ``$age$ must be used''

\item[feature importance measures] -- saying that certain features have greater importance than others or may have a rank order of importance.

\item[intermediate features] -- such as ``unusual behaviour'' that are used in the decision, but do not have an {\textit a priori} definition or derivation from the features.

\item[free text] -- which may introduce new terms, concepts or even rules, but embedded in unstructured text so that even identifying them is an issue.

\end{description}

Each of these create new algorithmic challenges for machine learning techniques.  In this section, we will look at each and suggest how they might be incorporated into specific ML approaches, but it is meant to be a starting point, not a solution.

As noted earlier, human explanations may be based on more or less expertise, and thus need to be weighed more or less strongly. 
In addition, some explanations concern a single data items whereas others may be counterfactual explaining different classification of two apparently similar examples.  Finally, the way these are incorporated into a model might be my (re)starting machine learning from scratch, whereas others may use incremental learning, updating the existing model.  We have not dealt with these issues explicitly in the potential approaches below, but they each add to the algorithmic challenges.

\subsection{Incorporating feature constraints into decision trees}

The examples in both Section~\ref{sc:ui-il} and~\ref{sc:ui-qbb} gave rise to data examples with a list of features with some sort of value constraint on each feature.  these constraints may be exact (e.g. \verb|loan<10,000|, or \verb|has_PhD=true|), partial (e.g. \verb|nos_phds| is large, or simply knowing that some feature should be included.  In the latter cases, the exact values for the feature will have to be learnt.  More abstractly, we have a set of examples, and for each example X a set of local criteria A, B, etc., for which we would like to ensure that A, B, ... are part of the decision that gives rise to X.

In the case of a conjunctive rule, it  must be of the form: $A \land B \land ...$

\begin{figure}
  \centering
  \includegraphics[width=0.3\linewidth]{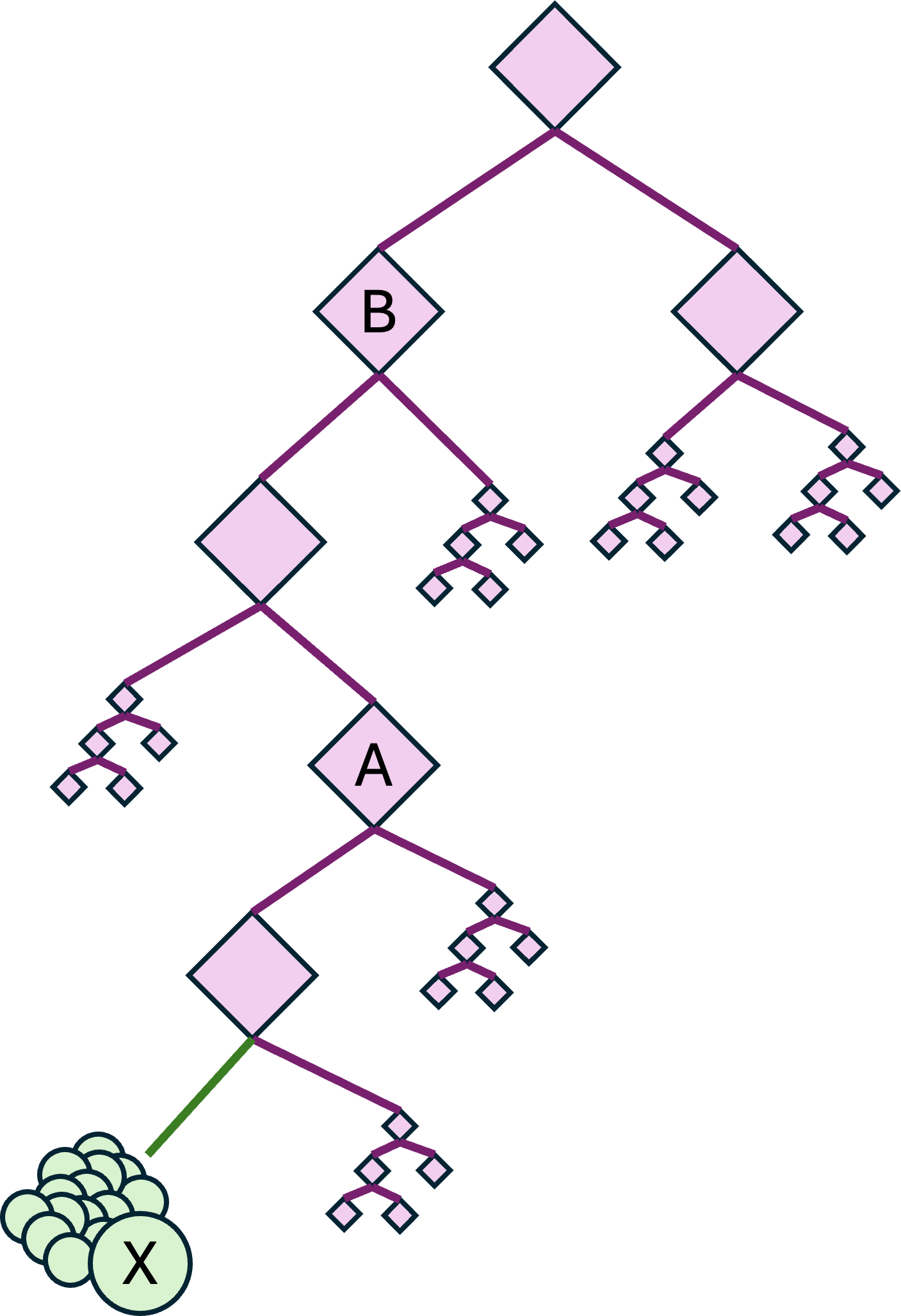}
  \caption{Constraining a decision tree to include explanation criteria.}
  \label{fg:constrain-dt}
\end{figure}

If we are generating a decision tree it must be constrained to include A and B above the node where X is classified (see Fig.~\ref{fg:constrain-dt}).

If the decision tree or rule set is generated by a GA or similar algorithms this can be added to the fitness function.

For top-down algorithms such as ID3 / C4.5 \cite{quinlan1986induction,quinlan2014c45} things are more complicated.  The criteria A, B, .. can simply be given additional weight alongside entropy, in a similar way to how one might use criteria from previous queries.  Alternatively one could start by forcing criterion A as the top node and B as second node of whichever branch X is classified.  The disadvantage of this is that this is effectively using them as global criteria, and hence hard to extend this to multiple examples X, Y, Z, with associated criteria.

Another option is to build a tree as normal and then force the key decisions back into the tree as modifications bottom-up from the leaf node where X is classified, pruning or rebuilding sub-trees.   One version of this would be to look for the closest ancestor node where there are both A and not-A examples, insert a new A criterion node just above and then rebuild the tree below.

\subsection{Incorporating feature importance measures into learning}

When users provide SHAP-style importance values for features influencing their decisions, these values represent a form of soft constraint on how the model should weigh different inputs. Unlike the hard constraints of decision trees where features must be present in specific positions, these importance measures suggest relative weights that should influence the learning process.

For neural networks, one approach is to modify the backpropagation algorithm itself to incorporate feature importance. The user-provided importance values can be propagated through the network alongside the regular activations, effectively modifying how strongly different parts of the network adapt to training examples. This is achieved by adjusting the learning rate during weight updates based on the propagated importance values, while ensuring that the standard learning behavior is preserved when no importance values are specified.

For gradient boosted trees, the splitting criteria could be modified to favor splits on features that users have identified as important. This is similar to how decision tree algorithms typically use information gain or Gini impurity, but with an additional weighting factor based on user-provided importance values. Features marked as highly important by users would receive a boost in their splitting score, making them more likely to appear earlier in the tree structure.

A key challenge in implementing these approaches is handling cases where different examples have contradictory importance values for the same features. One potential solution is to cluster examples based on their importance distributions, allowing the model to learn multiple valid approaches to the same problem. This would acknowledge that different experts might have different but equally valid reasoning patterns for reaching the same conclusions. Additionally, factoring in the level of expertise could help weigh each case, enabling the model to prioritize insights from more experienced or relevant sources.

\subsection{Intermediate level features and user generated vocabulary}

\begin{figure}
  \centering
  \includegraphics[width=0.8\linewidth]{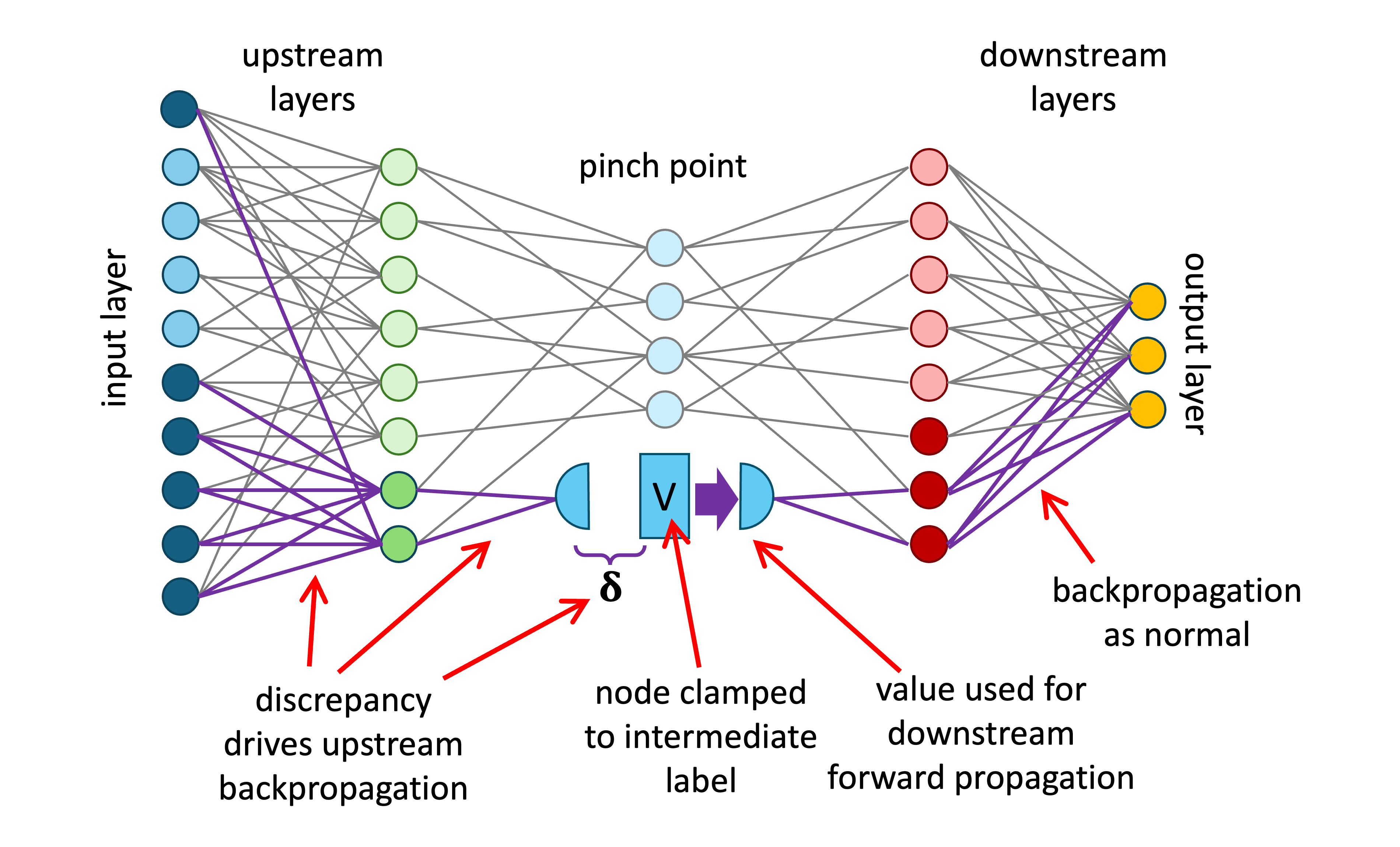}
  \caption{Clamping a deep neural net at feature-space level when semantic values known.}
  \label{fg:clamping-dnn}
\end{figure}

One way to deal with new intermediate level features algorithmically is to separately build a classifier for each new feature and then add these to the input features.  However, sometimes this additional information can be embedded more uniformly within the training process.  Figure~\ref{fg:clamping-dnn} shows one way to achieve this in a deep neural network.  An arbitrary node is chosen in a pinch-point layer, where one expects higher level concepts to be encoded.   When there is no coding for this intermediate feature backprop works as usual.  However, when the value is known the node is `clamped' to the known value.  The feedforward stage of backprop works as normal except the activation value for the node is stored and instead the known value is fed forward.  The feedback stage again works as usual, except that at the clamped node the difference between the known value and calculated activation is used as a delta value just as if it were on the output layer and then feeds backwards as normal.  The updates up and downstream of the clamped node may also be weighted more heavily.

\subsection{Free text explanations}

If free text user explanations emerge as part of interactions with LLMs, it may be possible to partially parse the explanation.  It might include existing feature names which can then be used to generate feature constraints or importance, and new terms might be introducing \emph{intermediate features}.  The explanations could include more complex logical arguments requiring deeper analysis, for example  such as ``although I wouldn't normally look at foot reflexes when dealing with kidney problems, because of the patients unusual medical history ...''.  This parsing may itself use LLMs or other NLP techniques. 

\note{Tom ... maybe ideas on how user explanations could be  re-embedded into prompts that say, ``when you see a situation like X bear Y in mind'' where X is a `training' example and Y the user explanation for X.  }

\section{Architectural implications}
\label{sc:arch}
\note{front-end to/from back-end API}
\note{WP4 = 'predict' ... fit ... explain()?}
\note{humanlayer.dev as existing example}

Supporting human explanations as part of AI system development requires rethinking traditional machine learning architectures. While traditional systems follow a linear pipeline from training to prediction, incorporating human explanations creates a cyclic flow where user feedback can influence both immediate responses and long-term model behavior.

Some existing frameworks have begun to tackle the challenge of integrating human feedback into AI systems. For example, HumanLayer\footnote{\url{https://humanlayer.dev}} provides an API for collecting human input at critical points in AI workflows. However, supporting rich explanatory feedback requires more comprehensive architectural changes.

The interactive nature of this architecture can be illustrated through the following pattern (where \mintinline{python}{clf} represents a generic classifier):

\begin{minted}{python}
# Simple one-way explanation
model_explanation = clf.explain()

# Interactive explanation with human feedback
with clf.explain() as model_explanation:
    # Initial model explanation available
    print(model_explanation)
    
    # Human provides explanation and gets updated model
    model_explanation = model_explanation.provide(human_explanation)
\end{minted}

\begin{figure}[!htb]
  \centering
  \includegraphics[width=0.8\linewidth]{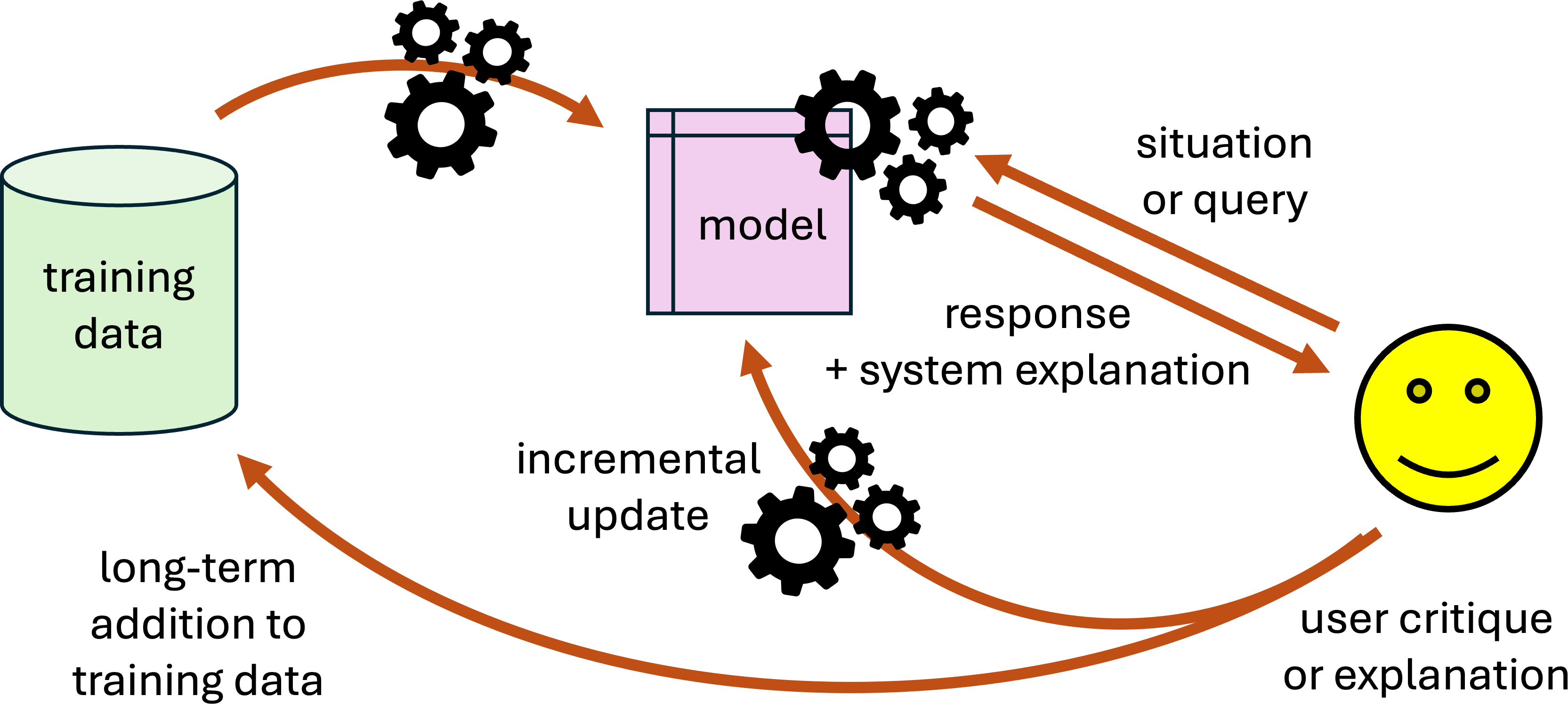}
  \caption{Architecture showing the dual feedback loops created by human explanations: immediate response updates and long-term training improvements.}
  \label{fg:arch-data-flow}
\end{figure}

This interaction creates two feedback loops, as shown in Figure~\ref{fg:arch-data-flow}:

\begin{itemize}
\item A short-term loop where human critiques and explanations can trigger incremental updates to improve immediate system responses
\item A long-term loop where validated explanations become part of the training data, influencing future model behavior
\end{itemize}

Supporting these loops requires several key architectural capabilities:

\begin{enumerate}
\item Interactive sessions that maintain context across multiple explanation exchanges
\item Version control for both models and their associated explanations
\item Mechanisms to incorporate human feedback into both immediate responses and training processes
\item Metrics to track how effectively the system learns from human explanations
\end{enumerate}

More complex forms of explanation, such as free-text rationales or intermediate features, would require additional processing components while maintaining these fundamental feedback loops. Integration with existing ML infrastructure platforms like MLFlow\footnote{\url{https://mlflow.org}} would need to support both the immediate interactive nature of explanations and their long-term incorporation into model development through appropriate versioning and artifact storage.

\note{different kinds of explanation from human to system: (i) data + explanation; (ii) counterfactual + explanatuon (may be differential explanation); (ii) system decision (+explanation) + user critique}

\section{Social implications}
\label{sc:social}
There are social implications of this sort of development that need to be considered, too. For human feedback to be effectively incorporated (whether in training or in active learning) the context of that feedback will need to be taken into account. Is it local feedback that applies in the local context? Or is it global feedback that applies universally? Is it expert feedback or novice feedback? The former likely to include a judgement on the accuracy and appropriateness of any AI explanation, the latter likely to cover only accessibility and user satisfaction in a broad sense. Who is able to determine this context? How is reliability or improvement generally evaluated?

More concretely, the staff at a given hospital -- though individually diverse -- will have a greater or lesser degree of common purpose according to their local health community and their own organisation's role within it. Similarly, an educational establishment will be attuned to the needs of their own students. In each of these example contexts, there are organic, social processes by which the sharing of developed resources beyond the original organisation can take account of the suitability of one context for the development of resources for use in another. So there are levels of meta information that may or may not become embedded in the technology -- dependent on not just 'what' but 'why' explanations. Grasping the meaning, significance and value of that meta information will be a new area for individuals and organisations. How will that be navigated? Who will control and curate the information?

Previous works \cite{greenhalgh2017AdoptionNewFramework,lehoux2017ProvidingValueNew} have drawn attention to the distinct interests and approaches of what they call upstream and downstream parts of the development and adoption process for technology. Is it likely that commercial, upstream stakeholders will prioritise the community interests of expert and lay humans who provide inputs for improved AI decision-making? If a user or group of users provides a huge amount of feedback that improves the AI learning and tailors an AI to their school, hospital, community etc, then the value of that trained AI may be great for similar groups. But if the process is owned and controlled by tech companies, then the ability to negotiate that benefit across communities of interest is lessened and might be eliminated altogether.

AI that can engage in meaningful dialogue on a specific problem may well become more persuasive. This is already a feature of LLMs where the form of the response can often appear more compelling than its content for the unwary. Bringing users closer into active dialogue may raise as many questions as it addresses.

\section{Discussion and Future Work}

\note{Short: reminder of key points; things we are planning (definitely incorporate some into QbB); open call for engagement. }

\note{Add notes here ta present: to inspire others, new ideas, ...}

\note{Include challenges for AI algorithms}

This paper has introduced the concept of human explanations as input to AI systems, moving beyond the traditional one-way flow of AI explanations to humans. We have explored several mechanisms for capturing these explanations, from feature constraints to free-text rationales, and proposed ways to incorporate them into machine learning systems.

Our immediate plans focus on extending Query-by-Browsing to implement human explanations as described in Section~\ref{sc:ui-qbb}. This will allow users to not just select examples, but also explain their selections through feature constraints and importance measures.

This work opens up several promising research directions:

\begin{itemize}

\item Developing algorithms that can effectively incorporate human explanations while maintaining model performance and generalization capabilities, particularly handling cases where explanations may conflict or require careful balance with patterns in the training data

\item Creating frameworks for translating different forms of human explanation -- from feature constraints to free text -- into actionable guidance for machine learning systems

\item Investigating how human explanations might help identify and mitigate algorithmic bias, potentially leading to more transparent and trustworthy AI systems

\item Extending these approaches beyond QbB to different types of machine learning models and application domains, exploring how various AI architectures might benefit from human explanatory input

\end{itemize}

The incorporation of human explanations into AI systems represents a significant shift in how we think about human-AI interaction. Rather than treating human input merely as training data, we propose considering it as a rich source of reasoning that can help shape how AI systems learn and explain their decisions. We hope this work inspires others to explore this promising direction for creating more effective and truly synergistic human-AI partnerships.

For ongoing updates on this work see \url{https://alandix.com/academic/papers/AXAI2025-talking-back/}.

\begin{acknowledgments}
  This work has been supported by the HORIZON Europe projects TANGO - Grant Agreement n. 101120763 and SoBigData++  Grant Agreement n. 871042.  Views and opinions expressed are however those of the author(s) only and do not necessarily reflect those of the European Union or the European Health and Digital Executive Agency (HaDEA). Neither the European Union nor the granting authority can be held responsible for them.

  This work was also produced with the co-funding of the European Union -- Next Generation EU, in the context of The National Recovery and Resilience Plan, Investment 1.5 Ecosystems of Innovation, Project Tuscany Health Ecosystem (THE), ECS00000017. Spoke 3.
\end{acknowledgments}

\bibliography{AXAI2025-talkback}

\end{document}